\documentclass[sigconf]{acmart}

\AtBeginDocument{%
  \providecommand\BibTeX{{%
    \normalfont B\kern-0.5em{\scshape i\kern-0.25em b}\kern-0.8em\TeX}}}

\copyrightyear{2024}
\acmYear{2024}
\setcopyright{rightsretained}
\acmConference[DIS '24]{Designing Interactive Systems Conference}{July 1--5, 2024}{IT University of Copenhagen, Denmark}
\acmBooktitle{Designing Interactive Systems Conference (DIS '24), July 1--5, 2024, IT University of Copenhagen, Denmark}\acmDOI{10.1145/3643834.3660714}
\acmISBN{979-8-4007-0583-0/24/07}

\begin{document}

\title{Ethics Pathways: A Design Activity for Reflecting on Ethics Engagement in HCI Research}
\renewcommand{\shorttitle}{Ethics Pathways}

\author{Inha Cha}
\authornote{Both authors contributed equally to this research.}
\email{icha9@gatech.edu}
\affiliation{ 
\institution{Georgia Institute of Technology}
\country{United States}
}

\author{Ajit G. Pillai}
\authornotemark[1]
\email{ajit.pillai@sydney.edu.au}
\affiliation{%
  \institution{The University of Sydney}
  \country{Australia}
}

\author{Richmond Wong}
\email{rwong34@gatech.edu}
\affiliation{ 
\institution{Georgia Institute of Technology}
\country{United States}
}

\renewcommand{\shortauthors}{Cha and Pillai et al.}

\begin{abstract}

This paper introduces \textit{\textbf{Ethics Pathways}}, a design activity aimed at understanding HCI and design researchers' ethics engagements and flows during their research process.  Despite a strong ethical commitment in these fields, challenges persist in grasping the complexity of researchers' engagement with ethics ---practices conducted to operationalize ethics---in situated institutional contexts. Ethics Pathways, developed through six playtesting sessions, offers a design approach to understanding the complexities of researchers' past ethics engagements in their work. This activity involves four main tasks: recalling ethical incidents; describing stakeholders involved in the situation; recounting their actions or speculative alternatives; and reflection and emotion walk-through. The paper reflects on the role of design decisions and facilitation strategies in achieving these goals. The design activity contributes to the discourse on ethical HCI research by conceptualizing ethics engagement as a part of ongoing research processing, highlighting connections between individual affective experiences, social interactions across power differences, and institutional goals. 

\end{abstract}




\begin{teaserfigure}
  \includegraphics[width=\textwidth]{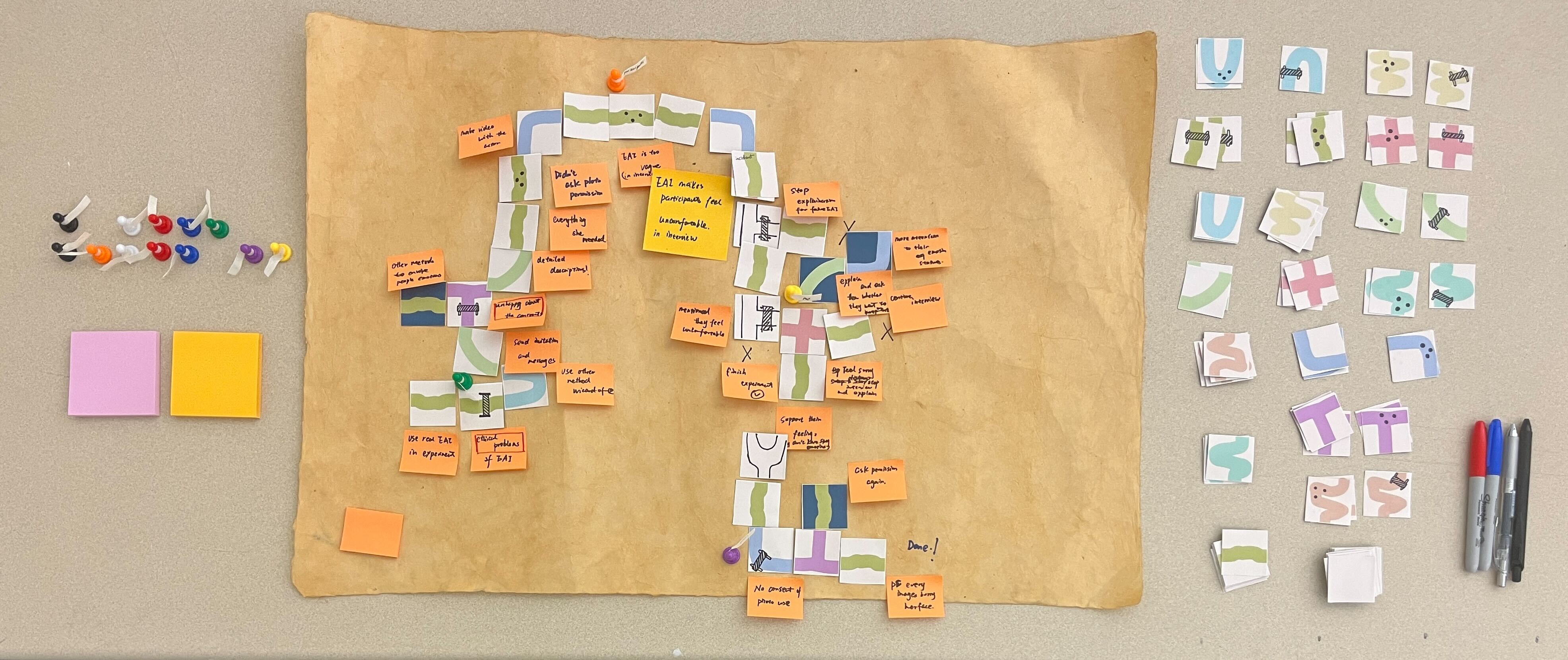}
  \caption{This example showcases the results of a participant and demonstrates the various elements and settings utilized in our activity, including a map for participants to create a set of paths depicting their engagement with ethics (center); a set of path cards to metaphorically represent different steps, decision points, and obstacles in their process (right); a set of pawn pieces to represent stakeholders in the story, and sticky notes to annotate the steps along the path (left).}
  \Description{In the center of the desk, there is a background paper serving as a space for participants to arrange their road cards, environmental elements, and sticky notes. This picture shows what a participant made through the activity. To the right, we have road cards for easy accessibility during the activity. To the left, we set colorful pawns opting for character forms with minimal inherent meaning. Characters and corresponding incidents or endings are represented through the placement of these pawns and sticky notes.}
  \label{fig:teaser}
\end{teaserfigure}


\maketitle

\section{Introduction}
Research ethics has been a cornerstone of Human-Computer Interaction (HCI) research for decades \cite{Mackay1995Ethics}, with courses \cite{munteanu2021dealing}, workshops \cite{g2021co}, SIGS \cite{fiesler2022sigchiethics}, and panels \cite{bjorn2018groupethics,Densmore2020research,fiesler2022research} being organized at major conferences in the field to help researchers in the field learn about and reflect on ethical issues in their work. 
Furthermore, there is a growing interest in navigating the evolving norms and expectations surrounding research ethics in the fields of HCI and design research \cite{Densmore2020research, brown2016fiveprovocations, vitak2016beyondbelmont, spiel2019gendersurveys} due to the complexity faced by researchers working with technology and new forms of data collection that have socio-technical implications, while also increasingly working in sensitive settings with marginalized communities. While HCI researchers typically undergo formal ethics reviews for their evaluation or field research protocols \cite{Munteanu2015situational} (such as through institutional review boards or other forms of industry ethical research review, e.g., \cite{cohn_jackman_molly_research_2016}), establishing a cohesive understanding of research ethics remains a persistent challenge within the HCI community. This challenge is further intensified by the diverse disciplinary traditions within the research community, the constant evolution of technology, research methods, and the variations in standards among different institutions and governments \cite{fiesler2022research}. One significant issue arises from potential discrepancies between the approved protocols by research ethics boards and the practicalities of conducting HCI research in nontraditional research environments outside of the "lab" \cite{Munteanu2015situational}. Relying solely on formal documentation as a means for ethical research may overlook the importance of researchers engaging in moral reflection and considering the ethical implications of their research itself \cite{Munteanu2015situational}. While ethics in HCI encompasses both research ethics and product design ethics, we focus on research given the rich discussion within HCI on researcher ethics, as well as our own positionality as researchers.

Shilton \cite{Shilton2013value} underscores the significance of recognizing the contextual nature of ethics, the researcher's role, and the influence of systems and infrastructure within the institution. In navigating the intricate interplay of ethics and research, it is crucial to acknowledge that HCI and design researchers are also inherently embedded within institutional contexts. They consistently interact with individuals, tools, resources, and administrative processes, collectively shaping the institutional infrastructure. The ethical dimensions associated with HCI practices are inherently complex, contingent, and intertwined with various factors, including individuals' ethics knowledge and judgment \cite{Gray2023scaffolding}. Despite HCI researchers' awareness of ethical issues, there exists a gap in understanding the ethical complexities they experience \textit{during} their work practices. 
Previous studies have explored how researchers navigate and respond to ethical challenges through conventional research methods such as interviews, surveys, or literature reviews (e.g.,\cite{vitak2016beyondbelmont,nunes2022scopingreview,salehzadeh2023changes}). 

In this paper, we take a design-oriented approach to understanding researchers' past ethics engagements in situated contexts. Ethics engagement involves taking steps to put an individual's ethical values into practice within a specific context. We view the collective aspect of ethics engagement as "ethics flows"—the dynamic and interconnected processes through which ethical considerations are navigated and managed within the research environment. 
We reflexively created a design activity as a vehicle for knowledge generation, both in the sense that (a) we thought about and learned how to conceptualize ethics flows and ethics engagement through our hands-on immersion in making the activity, and (b) the activity itself can be used with participants to helps us understand other people's practices.

We present our design activity, \textbf{\textit{Ethics Pathways}}, which aims to understand the nuanced process of navigating engagement with ethics throughout the research journey. We developed the activity and refined its mechanics across multiple pilot studies and six play-testing sessions. Play-testing participants used the activity to reflect on ethics engagements such as inadvertent data recipients, the moral complexities associated with probing into people's emotional states, and the moral quandaries faced by a junior researcher in upholding ethical standards during unethical corporate research and development projects. 
Ethics Pathways consists of four main steps: \textit{1) "Incident Description,"} where participants recall and explore the origin of an ethical issue; \textit{2) "Character Design,"} where participants recount the roles of relevant stakeholders by describing them as characters in their ethics narrative; \textit{3) "Path Design and  Action Notes,"} involving reflection on chosen paths, possible alternative actions, and barriers to action related to their ethics engagement by using a deck of cards we created; and finally, \textit{4) a "Reflection and Emotion Walkthrough,"} where participants revisit their narrative, addressing reflection questions from facilitators to delve into emotions, resources, and lessons learned in the ethical decision-making journey.

This paper provides two main contributions. First, it contributes the design activity of Ethics Pathways and documentation of the authors' reflections about what we learned during the design and facilitation process. Ethics pathways can serve as both a way to gain insight into people's engagement with ethics processes and to help activity participants self-reflect on their own practices. Second, through the design of Ethics Pathways, we contribute a conceptual framing of ethics engagement and flows that views research ethics as a set of ongoing practices that are affected by a combination of individuals' affective experiences, their social interactions across power differences, and the goals and politics of the institutions they work within. This framing opens up new opportunities for HCI and design research to think about research ethics beyond individual decision-making, to also consider interventions or tools that can address issues related to social and institutional power structures.

The paper begins by reviewing related works and detailing the development of the Ethics Pathways. It then presents the final version of the Ethics Pathways activity. Following this, the paper reflects on the lessons we, the authors learned through the changes in our design and facilitation practices, discusses how the activity's conceptual lens of ethics engagement can help us better understand ethics in situated practice, and consider future uses and applications for Ethics Pathways.  

\section{Related Work}

\textit{Ethics Pathways} builds upon various previous initiatives that employ complementary methods to explore and highlight discussions on values and ethics through design approaches.

\subsection{Conceptualizing Ethics Engagement as a Part of Practice}
Our design activity investigates \textbf{ethics engagement}---the array of individual actions taken to operationalize ethics. We use the term ethics engagement to refer to a collection of perspectives that views ethics as related to practice. We describe these perspectives and how we see them fitting together. 

This paper conceptualizes ethics as part of the everyday work practices of HCI and design researchers, building on a broader body of research that connects ethics to situated and lived experiences \cite{jafarinaimi2015values, ledantec2009values, gray2019ethicalmediation}. In using the term "ethics" we refer to moral debates about social values, politics, and power \cite{shilton2018values}. Following past work informed by practice-led approaches, we do not pre-define what gets included or excluded as an ethical issue, leaving that as an "actor's category," allowing research participants to define it as they understand it \cite{kuutti2014turn, stolterman2008nature}. However, there are several aspects of conceptualizing ethics that we find important. Because \textit{we view ethics as being a part of situated and lived experiences}, ethical practices are also situated in specific organizational and institutional contexts, and affected by things like organizational structure, organizational incentives, interactions among stakeholders, resources, and infrastructures, organizational policies, or social power dynamics \cite{gray2019ethicalmediation, chivukula2023wrangling, Chivukula2020dimensions, wong2021softresistance, ali2023walking, deng2023investigating}. Furthermore, because ethics are a part of lived experiences, navigating ethical dilemmas and making ethical decisions are also embodied, felt and affective processes \cite{garrett2023feltethics}.

We also draw on research that highlights critical reflection as an important part of professional design practice \cite{schon2017reflective,adam2021design}, which can also lead to reflection and consideration of the ethical and values-laden dimensions of one's work \cite{Sengers2005,friedman2013value, Munteanu2015situational}. Engaging in ethical reflection fosters a deep-seated understanding of moral values, enhancing one's ability to discern right from wrong in complex scenarios \cite{schmidt2014questioning, padiyath2024realist}. Over time, these reflective practices become intrinsic to the professional's character, guiding not just isolated decisions but shaping their entire approach to their research practice \cite{Chivukula2020dimensions, antes2012applying, singhapakdi1996moral}.  

We are also inspired by the concept of ethics flows, which emphasizes that ethical decision-making is influenced by: personal codes; informal workplace ethics; formal codes, policies, and frameworks within institutions; and the social infrastructure within the organization\cite{pierce1996computer}. Chivakula et. al. share this understanding of "ethical flow" by showing how the journey of an individual within the design process is an entanglement of the social values of an individual, the organization, and the situated context of the design process \cite{Chivukula2020dimensions}. In their work, they also highlight a gap in understanding 'ethical flows' and call for future work in this space \cite{Chivukula2020dimensions}. Our work seeks to address this by developing a design activity to enhance the understanding of HCI and design researchers' situated ethics engagement practices \cite{boyd2021adapting}. 

Ethics engagement thus consists of the array of individual actions taken to operationalize ethics within a situated context. Like ethics flow, we view ethics engagement as a dynamic process that interconnects individuals' decision-making and affective experiences, with the broader institutional infrastructures and social dynamics that they must interact with and operate within. Engagement with ethics also includes practices of critical reflection, enabling designers and researchers to assess situations ethically, take responsibility for engaging with ethical issues, and make informed decisions \cite{green2021contestation}. 
 

\subsection{Reflection on Research Ethics in HCI}

Research ethics in Human-Computer Interaction (HCI) is a critical area of concern, as it involves the responsible conduct of research and the consideration of the impact of HCI technologies on users and society, as evidenced by multiple panels and town halls at major HCI conferences (e.g. \cite{fiesler2022sigchiethics, Densmore2020research,bjorn2018groupethics,frauenberger2017ethicsHCItownhall}). Ethical considerations in HCI research encompass a wide range of issues, including user privacy, informed consent, data protection, and the potential for technology to influence user behavior and decision-making \cite{ahuja2021ethical}. In the domains of HCI and design research, there is a burgeoning interest in grappling with the shifting norms and expectations pertaining to research ethics. Although HCI researchers often undergo formal ethics reviews for their evaluation or field research procedures, fostering a unified comprehension of research ethics continues to be a persistent challenge within the diverse HCI community \cite{abbas2019ethics}. 

Formal ethics reviews for research (such as through institutional review boards) do not capture or address all ethical issues in research. Ethical issues and harms can still occur even if a research protocol is approved by an ethics review board, such as privacy issues that persist even when using publicly available data, which often only need a light ethics review process, if any, especially in a U.S. context \cite{vitak2016beyondbelmont,fiesler2018weareproduct}. Challenges posed by studying emerging technologies---such as artificial intelligence and big data---may not be adequately addressed by traditional ethics guidelines \cite{gray2018dark}. The disparate forms of labor conducted by researchers, as well as how researchers engage with project "failures" also have ethical implications for how researchers are accountable to one another and to their research subjects \cite{howell2021crackssuccess}. Conducting research with participants can present ethical challenges when unexpected situations arise during the research process, especially when working with sensitive issues or vulnerable groups \cite{waycott2015ethical}. Direct engagement with participants, such as in participatory design, raises specific ethical issues such as power dynamics, representation, and the potential for unintentional exploitation of participants \cite{Munteanu2015situational,parvin2018doing}.

These issues suggest the need for HCI and design researchers to critically reflect \cite{Sengers2005} on research ethics, beyond formal ethics review processes. For instance, Munteanu and colleagues advocate for ethical reflexivity in participatory HCI research, where researchers continuously reflect on and adjust their methods to ensure that the rights and well-being of participants are safeguarded throughout the research process \cite{Munteanu2015situational}. Others describe the need for continual and open reflection about ethical issues as a part of HCI and design scholarship \cite{nathan2016disruptions} and as a moral professional responsibility \cite{durgahee1997reflective}.

However, the ethical tensions experienced by HCI and design researchers within their situated work practices are not adequately explored. There is much prior work studying ethical tensions experienced by computing and design practitioners situated in product development practices (e.g., \cite{gray2019ethicalmediation,wong2021softresistance,pillai2022exploring, Chivukula2020dimensions,popova2024whoshouldact,widder2024power,deng2023investigating}). However when considering ethics in HCI and design research, most prior work seeks to understand ethical practices through the use of literature reviews \cite{salehzadeh2023changes,nunes2022scopingreview}, theoretical exploration \cite{brown2016fiveprovocations}, critical self-reflection on the authors' practices \cite{howell2021crackssuccess, Munteanu2015situational}, or surveys \cite{vitak2016beyondbelmont}. Few studies are seeking to study how other researchers approach ethical tensions in their work. Some of this may be in part to difficulties due to concerns about anonymity, sensitive details, or participant confidentiality. Nevertheless, this leads to a limited understanding of the considerations and steps that researchers take to resolve these tensions or ethical dilemmas. This gap highlights the need for more in-depth exploration and documentation of the ethical decision-making processes in research, which are crucial for comprehending and navigating the ethical landscape researchers encounter. 


\subsection{Ethics-Focused Design Methods}
A multitude of resources have been developed to facilitate engagement with ethical considerations in various contexts within the design process (as surveyed in \cite{chivukula2022surveying,friedman2017survey,wong2023seeing}). These resources include tools and methods that are primarily intended to comprehend user perspectives or consider how users and stakeholders may experience harm and range from simple checklists to more complex approaches. Examples of these approaches include speculating future scenarios \cite{wong2021timelines}, analyzing stakeholder dynamics \cite{friedman1996value}, and employing value-based methods \cite{friedman2013value, nathan2008envisioning}. Broadly, these methods tend to focus on creating more just and ethical \textit{outcomes} of (product) design processes and do not necessarily provide insight into the ethical dilemmas faced in \textit{research processes and practices}. 

Recently, the tech industry has seen a surge in the creation of tools, methods, and activities specifically designed for tech workers. Notable examples include Microsoft’s Inclusive Design Toolkit \cite{fraga2020inclusive} and IBM’s Everyday Ethics for AI \cite{rossieveryday}. Research studying how computing and design practitioners engage in ethical tensions and dilemmas in situated organizational practice has led to new design activities to help practitioners navigate these situations. For instance, Gray, Chivukula, and colleagues have created practitioner resources to find existing ethical design methods and activites \cite{Gray2023scaffolding}, as well as creating workshops that help practitioners develop their own plans of action for when they encounter ethical tensions in their work \cite{gray2022practitioner, chivukula2023wrangling}. However, these focus on design practitioners (often in product development contexts), rather than the ethical tensions and dilemmas of researchers.

Multiple scholars have used design fiction as a way to reflect on ethical aspects of research practice, including writing fictional paper abstracts to consider potential positive and negative directions in future research \cite{blythe2014researchfiction,baumer2014chi2039}, exploring the limits of institutional review boards in identifying potential ethical harms that could result from the use of publicly available data \cite{pater2022nohumans,fiesler2019ethicalconsiderations}, or more broadly considering the downstream or unintended harms that could arise from HCI and design research projects \cite{soden2019chi4evil,sturdee2021consequences,lindley2017implications}.

While these approaches are highly useful, they also emphasize a focus on the ethical impacts resulting from research outcomes. This approach overlooks a critical aspect: the practices, interactions, and affective experience of researchers grappling with ethical tensions \textit{during their research process}. Recognizing this gap, our objective is to take a design-oriented approach to create an activity in which researchers can participate that helps us understand their practices and experiences related to ethical tensions and ethical decision-making, and help them reflect on their experiences. 

\section{Design Process}
In this section, we provide a high-level overview of the design process of Ethics Pathways. In Section 4, we present and describe the activity steps in the final version of Ethics Pathways, and in Section 5, we reflect on what we learned through the process of designing and facilitating the activity.

\textit{Ethics Pathways} is an activity aimed at understanding an HCI or design researcher's reflections about the complex and messy process of ethics engagement during the research process.  Participants tell a story about an ethical issue they faced before by creating a representation of a path on a map that represents their different steps or decision points. The activity also provides prompts that help participants reflect on and discuss: their own emotional experiences; the role of other stakeholders in the situation; and the role of the broader organizational or institutional context where they are doing their research. The activity comprises an introduction before the main activity, followed by four key tasks:
 
\begin{enumerate}
    \item \textbf{Sensitizing Participants to Their Ethical Issues: }Several hours or days before starting the activity, participants are invited to reflect on ethical issues they may have encountered during their previous or current research experiences and what actions they have taken.
    \item \textbf{Incident Description: } At the start of the in-person activity, participants are prompted to recall an engagement with ethics and their first instance of recognizing, realizing, or identifying the ethical issue, delving into its origin and awareness, which kicks off the process of ethics engagement.
    \item \textbf{Character Design: } Participants choose tokens to represent stakeholders, or "characters," in their narrative of ethical action, including themselves. 
    \item \textbf{Path Design and Action Notes: } This step constitutes over half of the activity's time. Here, participants use a set of physical "path cards" to create a map that reflects either the path of actions they have taken or alternative paths that they speculate that they could have taken. Participants annotate their paths with "action notes" to detail their stories, incorporating elements such as resources they used, other stakeholders' actions and reactions, and other personal reflections.
    \item \textbf{Reflection and Emotion Walkthrough: }Once their fictional map is complete, participants revisit their narrative and address reflection questions and prompts provided by facilitators. These questions cover a spectrum of aspects, including reflecting on emotions, resources utilized, and lessons learned throughout the ethical decision-making journey.

\end{enumerate}

In the remainder of this section, we provide more details about our process of developing Ethics Pathways.


\subsection{Motivation and Goals}
\label{sec:goals}
Our primary goal in developing Ethics Pathways was to craft an activity that could help us understand participants' experiences engaging with ethics and decision-making in their research. Specifically, we sought to understand how participants reflect on their previous encounters with ethical issues in their research, the actions they took, and the broader context of these decisions.

Our focus on action and decision-making draws on research that conceptualizes ethics and values as inherently being a part of lived experiences and human actions \cite{frauenberger2016inaction,ledantec2009values,jafarinaimi2015values}. A body of research extends this conception of ethics as situation-dependent and lived experience to focus on the emotional, affective, and felt aspects of engaging in ethical action \cite{su2021criticalaffects,garrett2023feltethics,popova2022vulnerability}. In order to surface researchers' experiences engaging with ethical issues, we are inspired by reflective design's call for using design to help support users' critical reflection that can "highlight the choices one makes in everyday activities and to offer up new choices that may not have been in the user's awareness" \cite{Sengers2005}---in our case, we focus on HCI and design researchers making these reflections. Given the personal nature of ethical issues, we are attentive to the fact that some participants may feel that these reflections are personal and sensitive. 

We also recognize that ethical action and decision-making are not solely based on individual actions, but occur within complex assemblages, including societal norms, institutional contexts, and organizational policies \cite{Shilton2013value, shilton2018engaging}. Much work has shown how ethical action in industry research and design is affected by organizational factors including social power, relations among different stakeholders, institutional infrastructures and resources, and different incentives that institutions and individual researchers might have \cite{Chivukula2020dimensions, wong2021softresistance, gray2019ethicalmediation,ali2023walking, widder2024power, popova2024whoshouldact, madaio2020codesigning}. Our design goal of Ethics Pathways is thus not only to facilitate a deeper understanding of an individual's own stance on ethical matters, but to also shed light on the interactions between individuals, their environment, and institutional structures. 


Specifically, our design goals for Ethics Pathways are to empower participants to:
\begin{enumerate}
    \item Engage in a thorough, step-by-step reflection of past ethical issues, enhancing their understanding and insight;
    \item Share their lived and emotional narratives, along with their preferred level of obfuscation when designing characters or describing stories;
    \item Map and analyze the institutional infrastructure that played a role in their ethical experiences by fostering a sense of embodiment in their recounting;
    \item Envision alternative actions and decisions they would have liked to take in an ideal situation;
    \item Discuss the complexities and interactions with other individuals and communities involved in addressing ethical issues.

\end{enumerate}

\subsection{Design Inspiration}

To create an activity to help researchers reflect in ways that would achieve the above goals, we considered various metaphors, narrative styles, and world-building mechanics to structure our activity.

The main metaphor we use in the activity is "paths." We initially explored a range of metaphors in our early design explorations, inspired by ethics flow to represent a complex ecosystem before settling on the metaphor of paths. The chosen metaphor, paths, not only captures the dynamic nature of ethical decision-making but also provides a familiar and versatile framework for building narratives. This approach allows for step-by-step consideration, accommodating the complexities of ethical experiences. We also explored various narrative styles and world-building mechanics. We delved into methods like \textit{'Timelines \cite{wong2021timelines}'} and \textit{'Ethics Pondscape \cite{pillai2022exploring}'}, as well as design approaches like 'user journey mapping' \cite{endmann2016user} and the hero's journey \cite{sarantou2018hero}, ultimately choosing pathways as our primary modality for its simplicity and ability to provoke deep reflection.

Our design process also incorporated insights from world-building mechanics, emphasizing path-building inspired by 'Dungeons and Dragons \cite{crigger2021exploration}.' Additionally, we drew from the interactive aspects of 'Saboteur \cite{saboteurgame} ' to enhance the design of our paths as cards. The integration of diverse narrative forms and game mechanics aimed to address multiple levels of experience, encompassing procedural nuances, roles, infrastructural dynamics, and emotional journeys. This approach seeks to facilitate a meaningful exploration of ethical issues, providing participants with the ability to reflect on both their individual experience, as well as the broader context of their decision-making 


Subsequently, we iterated on the design of the method multiple times, which we explain more in Section 3.4.

\subsection{Pilot Sessions and Playtesting Sessions}



\begin{table*}

\caption{Playtesting Session Participants}
\label{tab:participants}

\begin{tabular}{p{0.05\textwidth}p{0.15\textwidth}p{0.3\textwidth}
p{0.3\textwidth}}
\toprule
\emph{No.} & \emph{Current roles} & \emph{Research experience} & 
\emph{Research area}\\\midrule

P1 & PhD candidate and IRB student member & 3 Years academic research & 
Trust and adoption of technology  \\ 

\midrule

P2 & PhD Student & 1 Year academic and 9 years industry research & 
Tangible interfaces, Games, and VR technologies  \\
\midrule

P3 & Assistant professor & 7 Years academic research & 
Critical making, Emotional AI  \\
\midrule

P4 & PhD candidate & 3 years academic research, 7 years industry research & 
UX Research, Queer HCI, Tangible embodied interaction  \\
\midrule

P5 & PhD student & 2 Years academic research & 
Emotional AI  \\
\midrule

P6 & PhD candidate & 15 Years academic research with multiple institutions & 
Urban planning, Afrofuturism, media studies  \\
\midrule

\end{tabular}    
\end{table*}

We conducted pilot sessions and playtesting sessions with participants at a research university in the southeastern region of the United States between October 2023 and January 2024. We conducted informal pilot sessions \footnote{We distinguish between pilot sessions and playtesting sessions in our paper. During pilot sessions, we informally engaged participants in activities designed to assess the viability of our initial idea of activities, design artifacts, and query protocol. We did not collect data during pilot sessions, many of which involved colleagues who were already familiar with our research motivations. We refined our approach based on feedback from the pilot sessions. This iterative process helped us develop Ethics Pathways by considering participant feedback in conjunction with our original motivations and goals.} across three sessions with four participants to test and iterate early versions of our activity, to develop the basic structure and mechanics, leading to the development of the five steps as outlined in Section 3. 

We then conducted more formal playtesting sessions with six participants, where we recorded the sessions and asked participants for feedback in a more structured way. The purpose of the playtesting was to \textit{gather feedback and refine} our activity's designed elements (such as how to represent paths and characters) and facilitation procedures, rather than formally evaluating the activity. The protocol for playtesting sessions was approved by a university institutional review board. Playtesting sessions lasted between 50-90 minutes.

We purposively recruited participants from our university who self-identified as HCI or design researchers, and self-reported experiencing ethical issues or dilemmas while doing their research projects and have taken actions. Six people responded to our recruitment call, with a range of 2–15 years of research experience. Participants ranged from Ph.D. students in their first year to their final year and an assistant professor. (See Table \ref{tab:participants} for participant information.\footnote{ We did not ask participants for their gender but did ask them for their preferred pronouns to help facilitate our interactions with them. For anonymity, we do not share gender pronouns in the table, but report that 4 participants use she/her pronouns; 1 uses he/him pronouns; and 1 uses they/them pronouns.}) While we recruited participants from an academic setting, we purposively sought to include participants who also had industry research experiences, ending up with 2 (P2 and P4). One of these participants (P4) reflected on an incident that occurred while they were an industry researcher.

We requested feedback from each participant regarding the effectiveness of our method with a short semi-structured interview at the end of the playtesting session. This included asking about aspects that worked well and those that could be enhanced. Additionally, we asked participants whether they would consider using this activity in the future and under what circumstances. We were also interested in their opinions on the potential adoption of our method by others. Collecting this feedback helped us understand how our activity might work at facilitating reflections on ethical action and decision-making, and helped us identify areas for improvement. 

After each of the pilot sessions and playtesting sessions, the session facilitator(s) would share their notes and reflections from the session with all the authors about what worked well, what could be improved, and participants' feedback. Through these discussions, we iteratively refined the actions that participants take in each activity step, the physical design materials used in the activity, and our facilitation techniques and strategies.

\subsection{Iteration of Design and Facilitation}

During initial pilot sessions, we brainstormed a range of metaphors and activities that might help participants reflect on their experiences with ethical action and decision-making, including paths, flowing rivers, and ecosystems. We also considered having participants reflect on their past experiences by retelling their stories as metaphorical "superheroes" with different powers and weaknesses. The pilot sessions helped us identify that the "paths" metaphor worked well to help participants think through different past actions and decisions. After pilot sessions, we set out to create an activity that would let participants reflect on their past actions by representing them as a path; think about various characters or stakeholders involved in their experience; and consider the role of their institutional or organizational context.

We created the first version of Ethics Pathways for our playtest sessions, which we continued to iterate across our playtest sessions in response to participant feedback. In early playtest sessions, our focus on the "paths" metaphors led us to design an initial version of an activity where participants created a physical map of a complex landscape representing their ethical journey, seeking to immerse participants fully in the metaphorical world. Early iterations involved a set of printed cards representing different path steps, 3D-printed vehicles that could represent different characters and stakeholders that move along the path, and 3D-printed environmental elements to represent resources or contextual factors (such as a wall to symbolize feeling trapped, mountainous terrain to indicate the ups and downs, or thorns to represent difficulty within the path). The main goal of these sessions was to gather feedback and ideas on how to improve the various elements of the activity, rather than evaluating the content of participants' reflections and stories generated during the activity.

Based on playtesting feedback, several key insights emerged. Participants found the activity confusing due to the diversity of elements and the number of instruction steps we initially had to introduce each of the elements. 
To address this, we simplified the activity. Simplification was identified as a crucial step forward. Consequently, we removed aspects of the activity that involved using 3D-printed vehicles and removed the use of 3D-printed environmental elements, as these were deemed unnecessarily complex. We found it more effective to concentrate on path design in the activity, and to prompt participants to reflect on the role of other stakeholders and contextual resources through the facilitators' questions.

\begin{figure*}
    \centering
    \includegraphics[width=\textwidth]{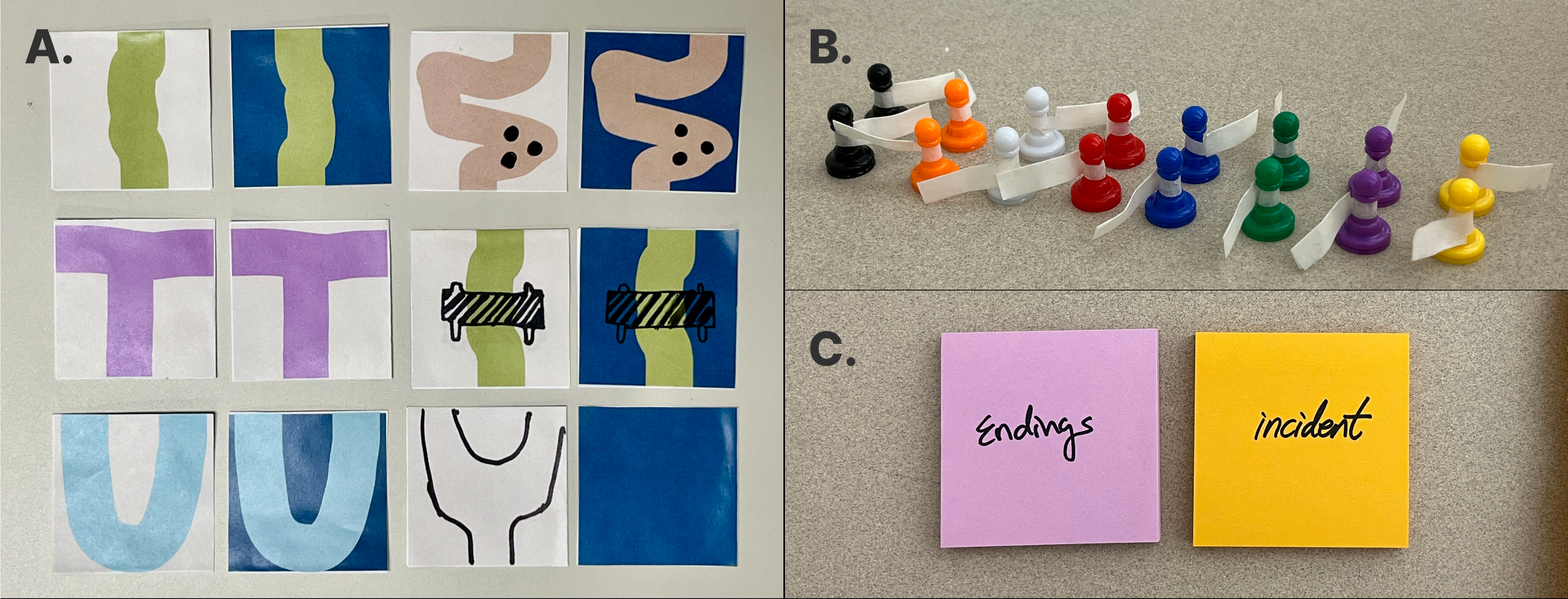}
    \caption{The final kit of materials for Ethics Pathways: A - new path card sets including a diverse set of paths in various shapes. The light and dark versions allow participants to distinguish between actual experiences and speculations about alternative paths that could have occurred; B - characters and stakeholders are now represented by using colorful pawn game pieces; C - providing sticky notes for participants to annotate their paths to describe incidents and endings.}
    \label{fig:finalversion}
\end{figure*}

With a renewed focus on the path design part of the activity, we found that participants were interested in exploring what could have been done in different situations had different decisions been made. To facilitate this, we redesigned our path cards to be double-sided(Fig. \ref{fig:finalversion}-A). We introduced different background colors to visually differentiate between actual experiences and alternative paths of action. This color variation allows participants to easily indicate whether their expressed thoughts represent the way they have chosen in the previous experience or speculative alternatives. 

Additionally, we shifted from using 3D-printed vehicle figures to more abstract and minimalistic character forms---a set of pawns--to help simplify the process of describing characters and stakeholders (as early playtesting participants spent a long time debating the metaphorical meaning of the different 3D-printed vehicles, which took away time from describing and reflecting on their past actions) (Fig. \ref{fig:finalversion}-B). While we simplified the number of steps in the overall activity, we also found it helpful to create a printed set of instructions and activity overview sheets to give to participants at the beginning of the main activities, helping them understand the whole activity before starting. 

In the next section, we will describe the specific details and step-by-step instructions for using our final version of Ethics Pathways.

\section{Introducing Ethics Pathways}
In this section, we present the steps of \textbf{\textit{Ethics Pathways}}. The main contributions of this paper are the proposed design activity itself, and our reflections on what we learned while developing and facilitating the activity. Analyzing the content of participants' narratives and experiences from the playtest sessions is beyond the scope of this paper; however, we provide illustrative examples of participants' stories throughout this section to illustrate the types of reflections that might occur during each step of Ethics Pathways. 

Each subsection begins by providing an overview of the facilitation instructions given to participants for each step and a description of the purpose of that particular step. Additionally, with their consent, we have included examples of our participants' works to share their experiences and further illustrate the process.

The activity consists of an introduction followed by four high-level steps, which can be summarized as follows:
\begin{itemize}
    \item Step 1: Sensitizing Participants to Their Ethical Issues (5 mins)
    \item Step 2: Incident Description (10 mins)
    \item Step 3: Character Design (15 mins)
    \item Step 4: Path Design and Action Notes (30 mins)
    \item Step 5: Reflection and Emotion Walk-through (15 mins)
\end{itemize}
The following materials are used during our activity (Fig \ref{fig:finalversion}):
\begin{itemize}
    \item Path Cards and Blank Cards (2in x 2in)
    \item Two Sizes of Sticky Notes (3in x 3in, 1.5in x 2in)
    \item Multicolor pawns
    \item Pens and Markers
    \item A0 Sized Background Paper (approximately 33in x 47 in)
\end{itemize}
Before starting the activity, we set these materials up on a desk. Each participant sat at the desk where materials were prepared. For our playtest sessions, we also included consent forms on the desk and set up a camera to record participants' actions. These are not inherently necessary to facilitate Ethics Pathways as an activity but can be useful if Ethics Pathways is being used as part of a research study.

\subsection{Step 1: Sensitizing participants to ethical issues}
We began to engage with participants before the activity session itself. 
Before the activity session, we emailed each participant, providing them with detailed information that included their session schedule, date, and location. This information was sent individually via email 3–7 days prior to the session. 
We also encouraged participants to reflect on any ethical issues they may have encountered during their previous or current research experiences and the actions they have taken to resolve them. 
We emphasized that if they have any questions or concerns, they should feel free to contact the researchers during this period.



Each session consisted of 1 participant and 1-2 facilitators. Once the participant arrived on the day of the activity session, we first asked them about their background information and demographics. The questions also included their previous research experience (e.g., research experience, types of research, organization types, and roles) and their exposure to ethical issues. 
By collecting this information, the facilitator can gain insights into the participant's backgrounds and experiences, which can help tailor the activity content and discussions accordingly. The participant and facilitator start to build rapport during this step. It also creates an opportunity for the participant to reflect on their own research experiences and any engagement with ethics.

We provided the participant with a single sheet of instructions outlining an overview of the main activity. This guided them to understand the various activities included in the entire session. Throughout the activity, we encouraged them to think aloud, sharing their thoughts and feelings as they participated.

\subsection{Step 2: Incident Description}

\begin{figure}
    \centering
    \includegraphics[width=0.5\textwidth]{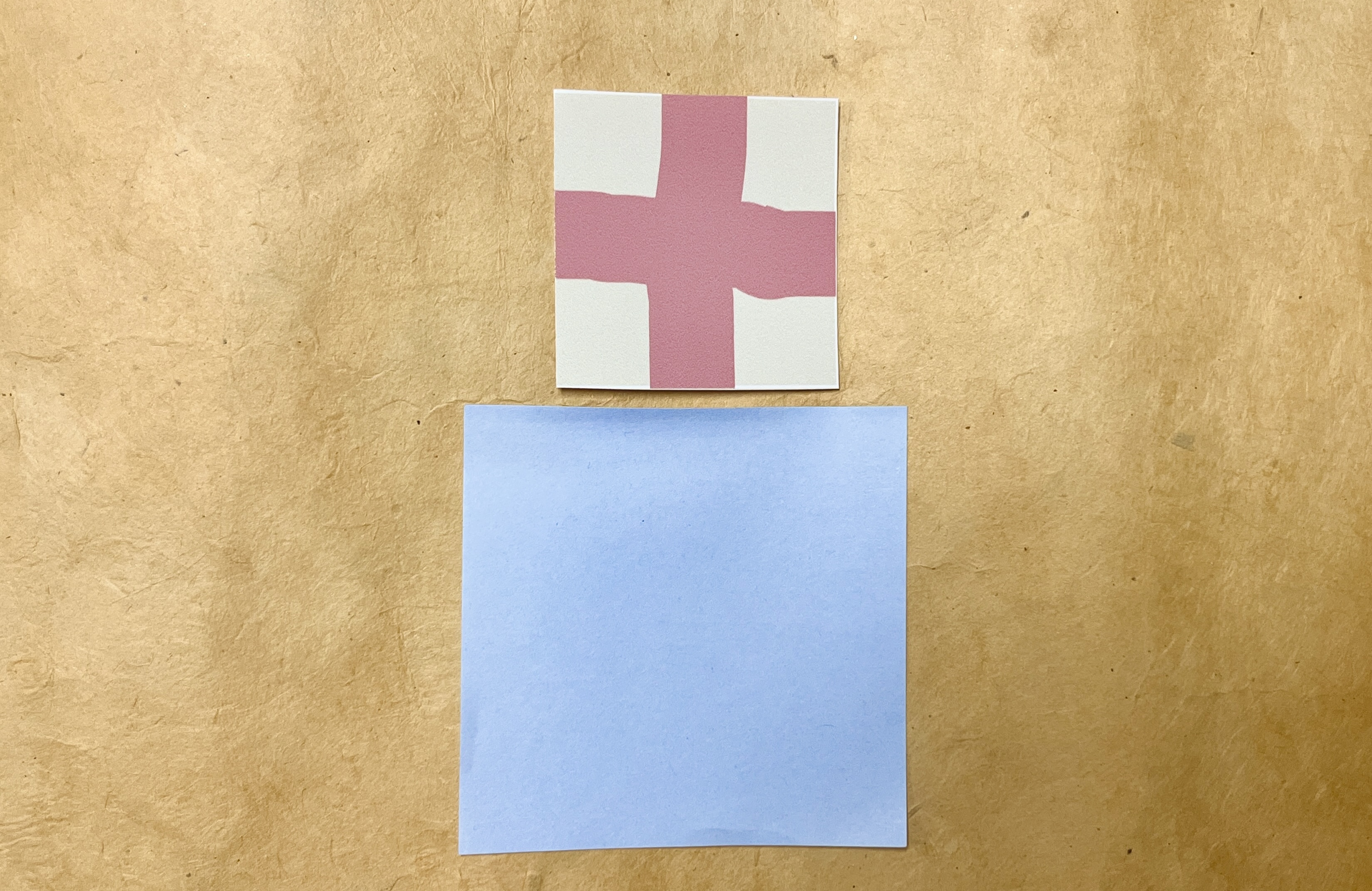}
    \caption{We provide the intersection path card with a blank sticky note to describe the incident at the beginning of the main activity. We inform them to write down the initial moments of the story they bring up during the activity.}
    \label{fig:intersection}
\end{figure}

To begin the activity, we placed the intersection path card (Fig. \ref{fig:intersection}) in the center of the map. We asked the participant to share their interpretation of the intersection and use it as a starting point to tell their story. The facilitator told the participant:

\begin{quote}
    Facilitator: “Now, let's take a moment to delve into the ethical issue you've encountered. Please share with me the initial interaction you had with the ethical issue. How did you first come across it? Reflect on the origin of the issue and consider how you became aware of it. Once you have formed an understanding, please write it down next to the square sticky note provided beside the path card in the middle of the map."
\end{quote}

 We allowed the participant to describe the incident as they wanted. If the participant wanted to share their experience in specific detail, they could describe it on the incident sticky note, or if they preferred to provide a more abstract representation of the event due to privacy and confidentiality concerns, they could describe it in a more obfuscated way.

\begin{figure}
    \centering
    \includegraphics[width=0.5\textwidth]{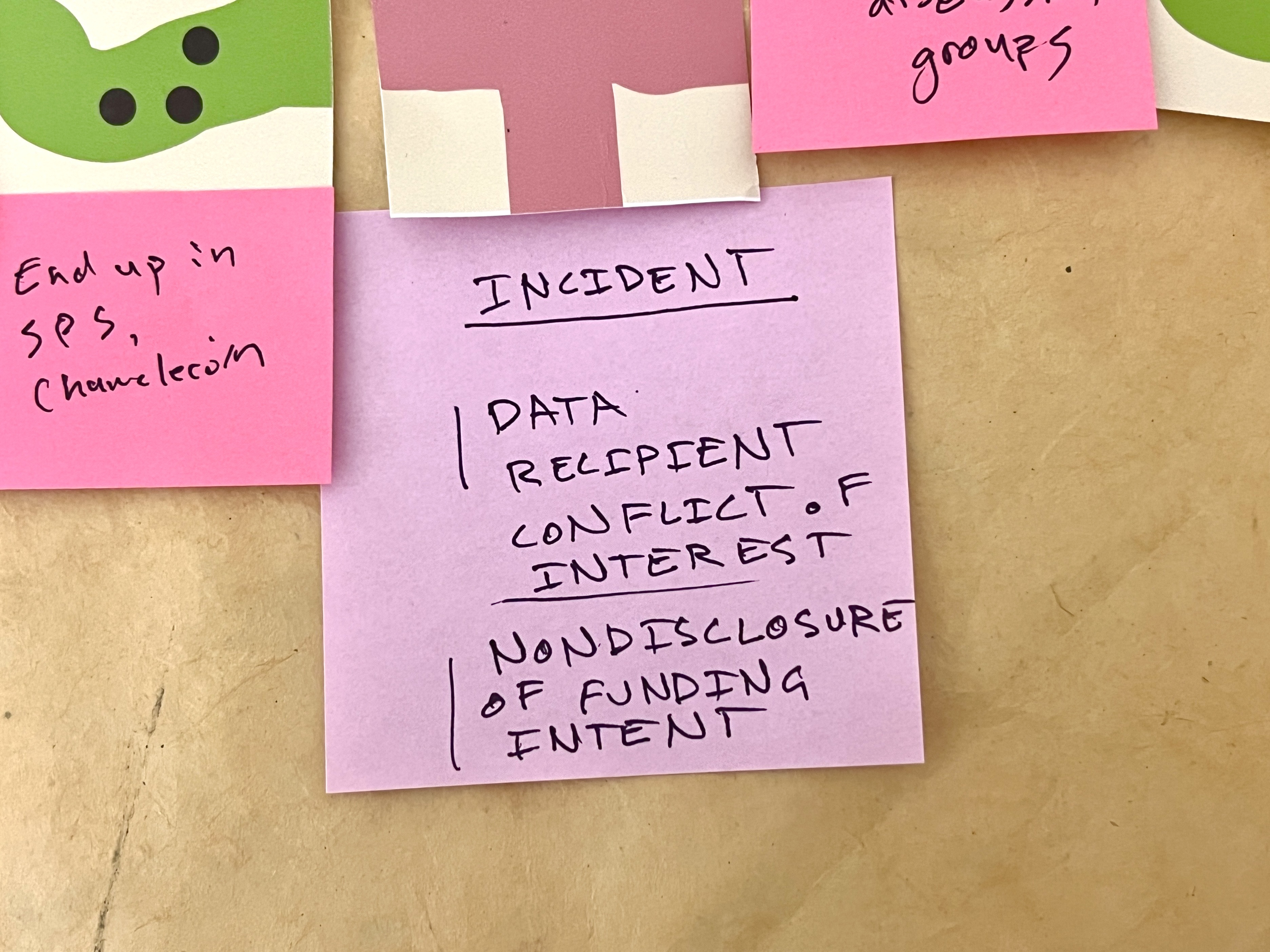}
    \caption{Example of incident card created by P2 in step 3}
    \label{fig:incident}
\end{figure}

\subsubsection{Example} P2 recounted a story that highlighted the ethical dilemmas associated with data collection and its ultimate recipients related to a past project. In this particular incident (Fig. \ref{fig:incident}), P2 and his colleagues collected data without being fully aware of its intended use, which turned out to be for the police and soldiers, which P2 felt raised ethical concerns. The lack of clear communication regarding the original intent of the project's funding contributed to this issue.

\subsection{Step 3: Character Design}
In this step, the participant was instructed to design characters representing themselves and other stakeholders in their stories. They were provided with a set of multicolored pawn game pieces which they could label. 
These labels could be identifying names or pseudonyms (again, allowing participants to obfuscate their stories in case of privacy concerns). We also encouraged our participants to provide additional information about the characters, such as their roles or associations (e.g., advisor, manager, participant, etc.).

\subsubsection{Example} 
\begin{figure*}
    \centering
    \includegraphics[width=\textwidth]{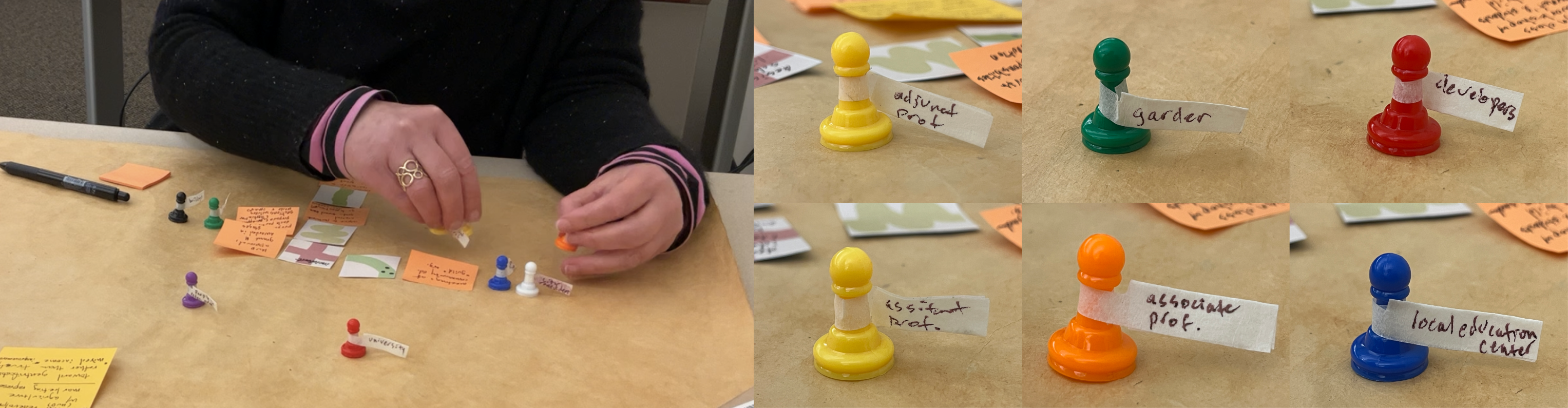}
    \caption{Left: P6 moving character pawns while explaining her story; Right: examples of characters designed by P6 representing stakeholders, including an adjunct professor, local gardener, developers, assistant professor, associate professor, and a local education center.}
    \label{fig:character}
\end{figure*}

P6 addressed ethical dilemmas in the context of redevelopment and urban planning projects. Their story featured different stakeholders (Fig. \ref{fig:character}), including three professors with different academic positions (assistant, adjunct, and associate). The project involved collaborating with the city's development department and local centers, and the characters in the story reflected not only individuals involved in the design process but also the institutes and organizations that took part in the decision-making journey.

\subsection{Step 4: Path Design and Action Notes}

After selecting characters in their stories, the participant was instructed to reflect on their actions and the role of characters in each step of their journey. This reflection was prompted by the use of two design elements: path cards and action notes.


The participants were provided with \textbf{path cards} that represent the steps of their story as it unfolds from beginning to end. Path cards were printed 2in x 2in cards that we created, which depicted various types of roads including straight roads, intersections, U-turns, winding roads, and blocks (Fig \ref{fig:finalversion}-A). They were encouraged to express their feelings and experiences through the shape of the road and other details. They could also draw or write their own different terrains or elements on the path cards to represent complexity, such as obstructions like dead ends, stop signs, or potholes to represent distractions. Light and dark versions of the path cards allow participants to distinguish between their actual experiences and speculations about alternative paths that could have occurred.

Additionally, we provided participants with \textbf{action notes}, small-sized 1.5in x 2in sticky notes, which allowed participants to annotate each path step to describe their actions, others' reactions, or the results of their actions. Participants were also asked to affix one action note for each corresponding path card. 

Combined, the path cards and action notes allow the participant to visually and narratively express their thoughts and experiences while reflecting on their journey. The facilitator introduced these elements by saying:

\begin{quote}
    
Facilitator: "While designing the path, you can describe the alternative path as well. For example, you can describe a situation in which you were aware of a possible action but couldn't execute it at the time. You can also describe a new path or action you discovered through reflection after the situation or while designing the map today. To represent alternative paths or actions that didn't happen in real-life experiences, you can flip the card and use a card with a different background color but the same shape."

\end{quote}

We concluded the path design exercise by requesting the participants to describe the endings of their stories and the factors that influenced the endings and to write those down on a sticky note and place them on their map as well.

\begin{figure*}
    \centering
    \includegraphics[width=\textwidth]{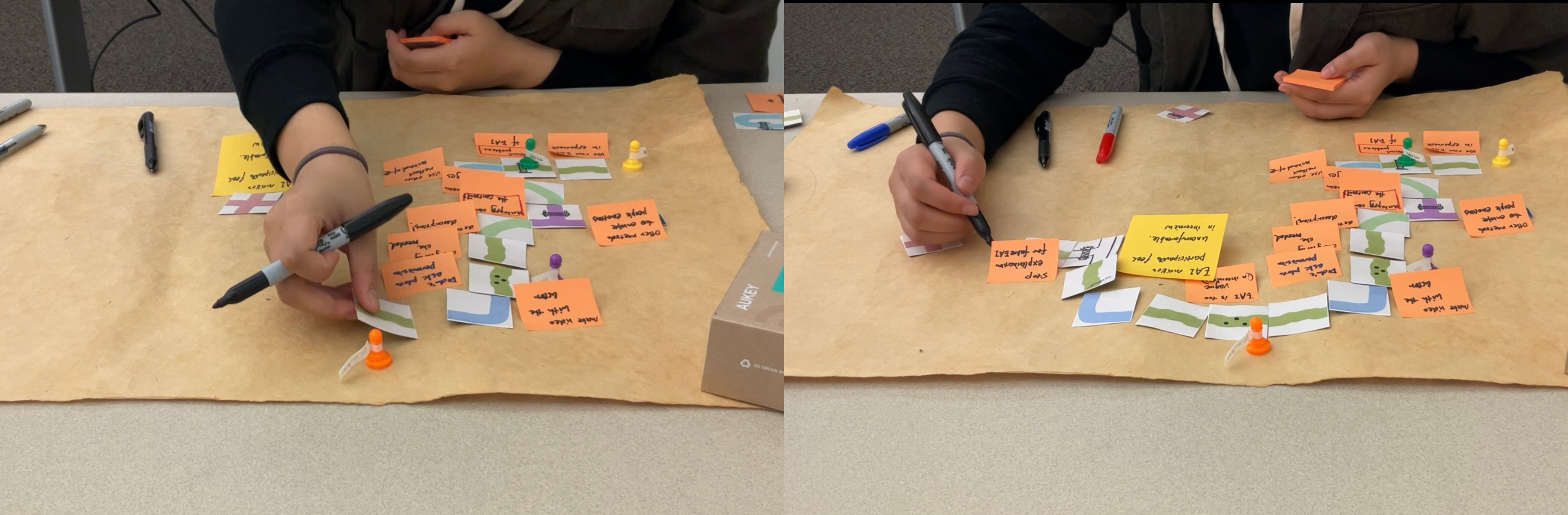}
    \caption{P5 designing her paths and adding action notes in Step 4}
    \label{fig:capturingthemoment}
\end{figure*}

\subsubsection{Example} 
\begin{figure*}
    \centering
    \includegraphics[width=\textwidth]{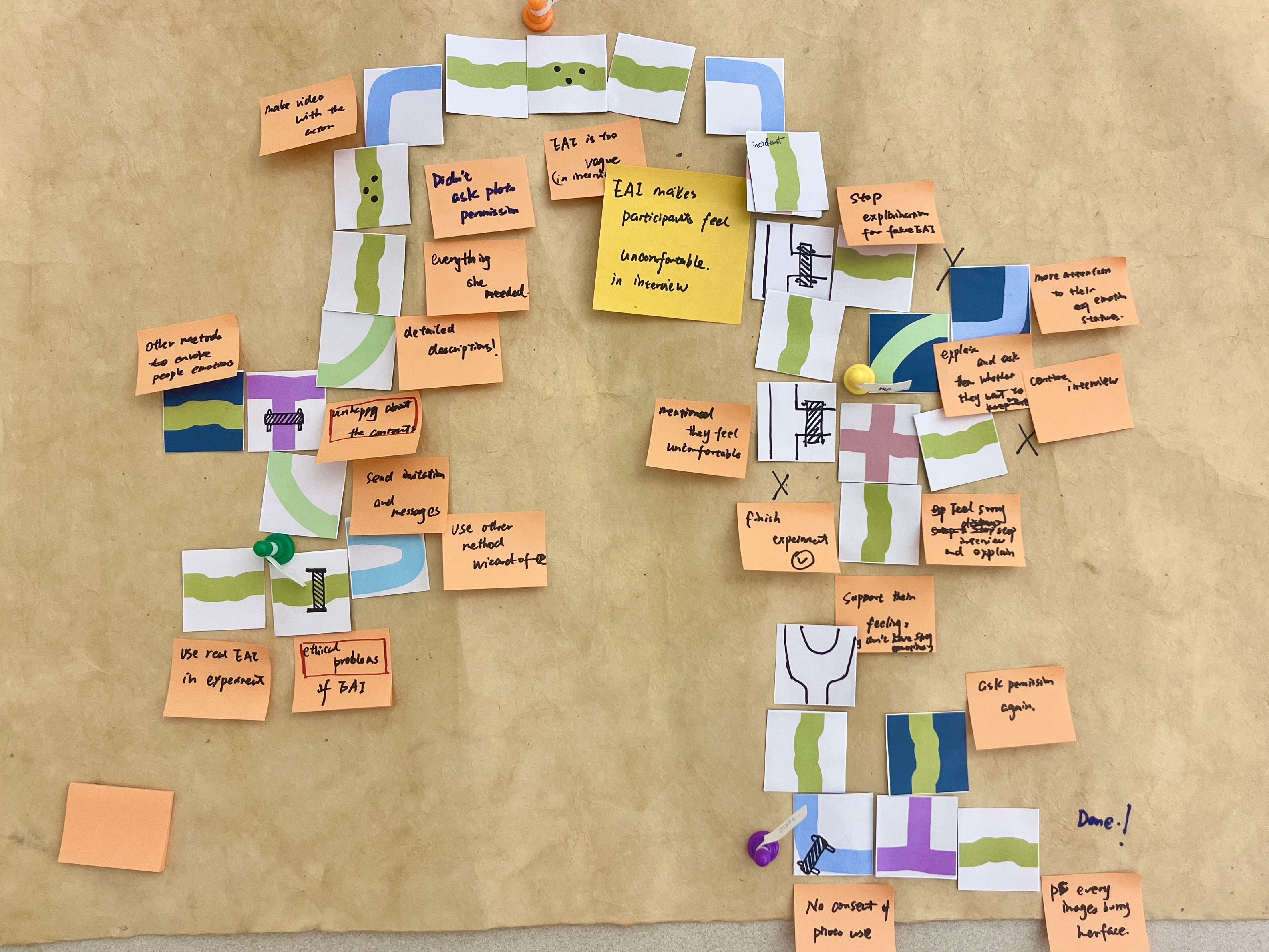}
    \caption{P5's session results}
    \label{fig:pathdesign}
\end{figure*}

P5's paths (Fig. \ref{fig:capturingthemoment}) revolved around the judgment and thought process regarding the issues that arose from the lack of sufficient prior notification to her research participants about the potential emotional discomfort they may experience during their involvement in a research project on emotional AI (EAI). She depicted the actions that caused the problems using path cards that had roads with holes, representing her mistakes. She used path cards that had blocked roads to portray actions that couldn't be taken due to opposition or the input of participants. For example, she wanted to use a real EAI system, but due to ethical concerns related to data training, her advisor made her use a wizard-of-oz EAI system. So, she placed a path card with a blocked road to symbolize an obstacle preventing her from using the real EAI system. Additionally, while reflecting, she depicted alternative paths by flipping the cards, indicating the actions she would have taken if given the chance. For example, she learned from this experience that users find it difficult to request pauses or share their emotions with researchers, even when experiencing emotional frustration (Fig. \ref{fig:pathdesign}). She emphasized the importance of researchers paying more attention to understanding users' emotions while participating in her study.

\subsection{Step 5: Reflection and Emotion Walk-through}
After completing the path design, we engaged the participant in revisiting their maps of their journey. By inviting participants to walk through their stories once again, we sought to gather a more detailed and nuanced understanding of their experiences. We encouraged them to recount their stories and provide descriptions of the stakeholders involved at different points along their journey. To capture a comprehensive understanding of their journey, we posed reflection questions to the participants. These questions covered a wide range of aspects, including stakeholders, emotions, resources, communities, barriers, changes, and lessons learned. Here are example questions we used in step 5: 

\begin{itemize}
    \item What were people’s reactions along the way?
    \item Were there any stakeholder concerns or feedback that influenced the project?
    \item What resources did you have along the way?
    \item If you could change one thing during the journey, what would you choose?
    \item What are the key takeaways or lessons you've gained from this experience?
\end{itemize}

We also prompted the participant to consider whose perspective should be taken into account when determining the success of the journey. In cases where multiple paths were possible, the participant was encouraged to describe each potential ending. Furthermore, if the outcome was portrayed as positive or negative,  the participant was asked to elaborate on why they felt that way, to gain a deeper understanding of their journeys and the various factors that influenced their ultimate outcomes.



\begin{figure*}
  \centering
   \includegraphics[width=\textwidth]{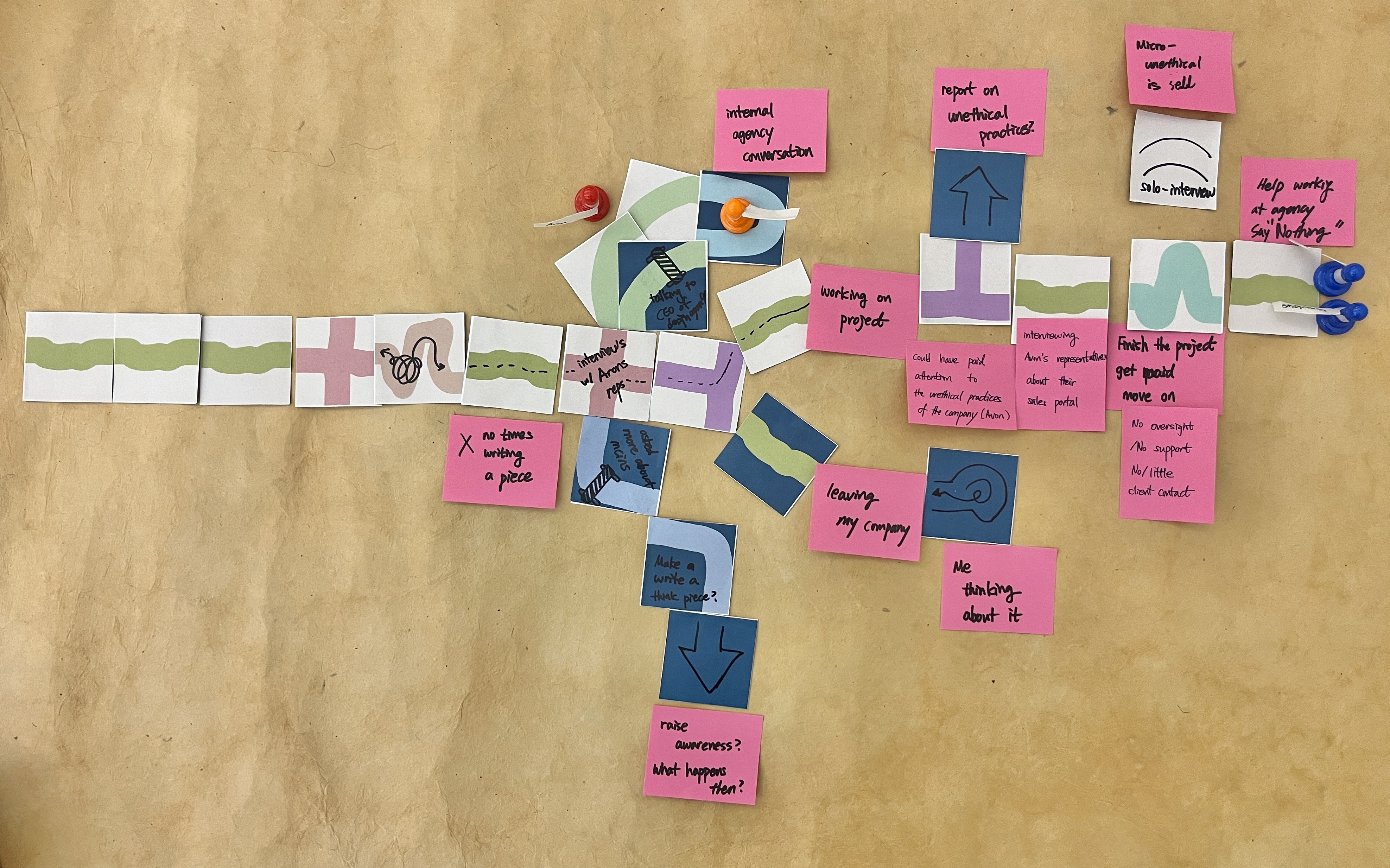}
 \caption{P4 session's results with three paths, including one path they went through before and two alternative paths. We replicated their map using our finalized kit, and the composition of our map was identical to their original version.}
    \label{fig:reflection2}
\end{figure*}

\subsubsection{Example} P4 described an ethical dilemma they faced when they were responsible for UX research for a multi-level marketing company at the design agency they were a part of (Fig. \ref{fig:reflection2}). They discussed the ethical issue of conducting research for a company that was not ethically sound. However, they could not refuse the task and continued with it, which led to problems. During the workshop, they designed three possible paths to address this dilemma. One path involved discussing the actual events that took place. A second path explored the actions they could have taken at that time but didn't. The third path showed new possible ways that they could consider during the session's path design, which they hadn't known about before. P4 mentioned the difficulty of raising the issue due to being in an entry-level position at the time and the lack of sufficient organizational support. They also discussed the multi-faceted nature of the organizational decision-making structure as a contributing factor to the issue. Reflecting on the activity as a whole, P4 said, “Oh, it just happened? I think that when I recall it, I was there were no decisions that I could have made. Like I literally didn't even think that I could build a map out of this issue because I thought well it just did the project and I didn't. (...) So I mean, maybe there's some way to differentiate in the cards like potential avenues versus actual avenues. I realized these were things that I could have done. But I didn't do. And having some cards built in to talk about things that were possible like on reflection like what was possible but was maybe closed off to me for whatever reason I think would be really helpful. (....) I like the challenge of having to actually think through the map because it made me realize that there were these possibilities, that there were these avenues and it made me more. Reflective of what was available to me, or maybe what was not available to me." Through a walk-through, they were able to think about the issue from multiple perspectives and consider the possibilities of alternate actions.



\section{Discussion}
We have introduced a structured, step-by-step approach designed specifically for HCI and design researchers to thoroughly examine and reflect on their engagement with ethics during their research experiences. In this section, we discuss three themes: 1) our self-reflections on how our design process and facilitation practices helped us achieve our design goals; 2) how we can understand ethical actions in practice through the lens of ethics engagement; and 3) how we consider Ethics Pathways in relation to other ethical reflection and speculation activities and consider future ways to adapt and use Ethics Pathways.

\subsection{Reflections on our Design Process and Facilitation Practices}
As described in Section \ref{sec:goals}, we described a set of goals that we wanted Ethics Pathways to help scaffold several actions for participants:

\begin{itemize}
    \item Engage in a thorough, step-by-step reflection of past ethical issues, enhancing their understanding and insight;
    \item Share their lived and emotional narratives, along with their preferred level of obfuscation when designing characters or describing stories;
    \item Map and analyze the institutional infrastructure that played a role in their ethical experiences by fostering a sense of embodiment in their recounting;
    \item Envision alternative actions and decisions they would have liked to take in an ideal situation;
    \item Discuss the complexities and interactions with other individuals and communities involved in addressing ethical issues.

\end{itemize}

In this subsection, we discuss how choices we made during our iterative design process and how choices we made in our facilitation of the activity collectively support these goals. 
Additionally, we consider how design and facilitation choices supported participants in anonymizing or obfuscating potentially identifiable details throughout the activity. 

\subsubsection{Achieving our Design Goals through Choices in Our Design Iterations and Facilitation Practices}

Our high-level goal was to provide an activity that would scaffold participants' ability to describe and reflect on their past experiences with ethics engagement. There are many elements that we wanted participants to describe and reflect on (including step-by-step actions, lived and emotional experiences, institutional infrastructure, envisioning alternate actions, and discussing interactions with other stakeholders). While at first, we considered creating different mini-activities or steps to focus on each of these elements separately, our iterative design and playtesting process led us to focus the final version of Ethics Pathways on the path design steps and integrate consideration of these elements into the ways participants did their path design.

Initially, our focus in character design (Step 3) was on having participants meticulously define the attributes, abilities, and traits of the central characters in their stories, to create fleshed-out personas that could be represented by 3D-printed vehicles. However, upon reflection among the authors and the realization that early participants spent a lot of time focusing on character traits, we shifted our emphasis. Instead of solely examining the competencies and individual traits of stakeholders, we directed our attention toward understanding the roles and interactions \textit{among} these stakeholders within their pathways. This led us to represent characters as pawn pieces, with labels that could be used to indicate their roles and could be easily "moved" along the path to depict the dynamic social interactions in participants' narratives. This allowed us to surface how these social interactions influenced participants' experiences of engaging with ethics. 

Considering the path design part of the activity, in early iterations of our activity, we initially assumed a linear progression in participants' stories of ethics engagement, with a clear starting and ending point. However, feedback from early pilot participants challenged this assumption. Some struggled to identify a definitive endpoint or resolution, while others described parallel activities leading to diverging paths. This prompted us to realize that our approach needed to support less linear explorations of participants' journeys. Consequently, in the final version, the path begins at the center of the map (Step 2) and can branch out in multiple directions without a predefined endpoint. 

Additionally, participants expressed interest in exploring alternative decisions when designing their paths or when doing the walk-through (Steps 4 and 5). We recognized the value of this approach for fostering critical reflection, by allowing participants to consider these alternatives or to help them reason through why they made the choices they did on their actual path. Consequently, in the final version of the activity, we explicitly provided support for participants to speculate about alternate paths using dark path cards and verbally suggesting to participants that they could depict alternate paths if they wished to do so. This allowance for envisioning alternative actions and decisions facilitated participants' critical reflection process, enabling them to articulate desired actions and recognize the barriers and constraints affecting their choices.

We also refined our facilitation of the study steps. One aspect we considered at the beginning of our design process was whether to incorporate think-aloud during the study. We debated between having participants explain their thoughts after completing the entire path or having them think aloud while designing each step of the path. Based on pilot sessions, we found that incorporating think-aloud naturally would be more effective in understanding participants' thought processes in their path design.

Reflecting on these decisions helped us think more about how we were conceptualizing ethics engagement and ethical flows, which we discuss more in Section \ref{sec:discussion-ethical-flows}.

\subsubsection{Fostering Confidential Discussions through Obfuscation, Abstraction, and Anonymization}

When discussing ethics within situated and sensitive contexts, it might be difficult for researchers to talk about prior ethical issues encountered in work due to confidentiality, privacy, IRB protocols, power imbalances among research collaborators, or other concerns. These might hinder open reflection and discussions about their experiences engaging with ethics. We consistently considered approaches to support open yet confidential discussions about ethics engagement. Hence, an approach that involves creating characters and using narratives, allows participants to describe their experiences at various levels of obfuscation based on their individual comfort. 

When creating characters, participants are asked to select a character pawn for themselves and label other pawn tokens to represent characters involved in the journey. These labels can be pseudonyms (as well as identifying names). The participants can assign different roles or associations to these characters at a level of detail of their choice, such as "manager" or any other relevant position. By utilizing characters to represent stakeholders and assigning them roles within the ethics journey, participants can explore the dynamics and relationships involved without revealing specific identities or organizations.

Similarly, with the use of path cards, participants have the flexibility to provide specific or abstract descriptions of their paths and actions. This flexibility allows participants to navigate confidentiality concerns by abstracting sensitive details while still conveying the essence of their ethics engagement. For instance, participants may use symbols like a blocked road path card to signify barriers encountered, while also being able to add depth by sharing the nature of the obstruction without disclosing identifying details about the specifics of that barrier.

Regarding facilitation, we also considered the physical location where we conducted the activity to safeguard confidential or sensitive information. In part, these choices were made in response to potential participants expressing concerns about confidentiality during recruitment. As part of these measures, we made sure that the sessions took place within closed environments and with one participant at a time to minimize exposure to others. Additionally, participants were assured of the confidentiality of their information, with explicit communication about the purpose of data usage and the exclusion of sensitive content from the dataset. During the activity, participants were given the option to designate certain information as "off the record" if they wished to exclude it from the analysis. These measures aimed to foster trust and confidence among participants, create a safe space where participants feel comfortable sharing their experiences, and enable the participants to engage in meaningful dialogue without apprehension about privacy concerns.

We next turn to discuss how our design and facilitation choices helped us critically reflect on our own conceptual lens of ethics engagement, and how that helped us think in new ways about ethical flows in practice. 

\subsection{Ethics Engagement as a Lens for Understanding Ethical Flows in Practice}
\label{sec:discussion-ethical-flows}
 
 To understand people's ethics engagements---the array of actions taken to operationalize ethics---we use our design activity to surface stories of ethics flows. Rather than focusing on \textit{outcomes} which is technical and design research often looks to when assessing ethics (such as identifying downstream harms and negative impacts) \cite{selbst2019fairnessabstraction, lindley2017implications}, we use the path activity to shift our focus to \textit{ethics flow}, which illustrates how individuals navigate ethical dilemmas intertwined with the social values of the individual, the organization, and the specific contextual environment. Interdisciplinary research drawing on legal and policy perspectives highlights that legitimate and ethical action is not solely dependent on outcomes, but also requires processes that are seen as ethical and legitimate \cite{selbst2019fairnessabstraction,freeman2000private,koops2008criteria}. By shifting our focus from outcomes to the flow, Ethics Pathways unravels the complexities of the research process, illuminating how individuals navigate their engagement with ethics. Highlighting what current practices and processes look like in organizational contexts and opening up discussion about what improved practices and processes might look like. Furthermore, some participants opt to omit an ending point of their ethical journey, or they arrive at the realization that a definitive solution may not exist. The design of the activity supports this, acknowledging and respecting the intricacies of non-linear narratives and recognizing that ethical dilemmas often do not follow a straightforward path and sometimes either do not have a resolution or the meaning of "resolved" is not shared between the stakeholders.


Through our activity, we observed that participants' ethical dilemmas are specific to their organizational and institutional settings. These factors play a significant role in shaping their ethics engagement and, eventually, the flow. To exemplify this, we share several example stories from our play-test participants:

\begin{itemize}
    \item P2 shared that his flow of ethics engagement was shaped by the complicated relationships between stakeholders that involved soldiers in the defense field, police, and funding institutions. With the pawn pieces set within the paths constructed by P2, they showcased their engagement with ethical tension caused by military personnel and police officers receiving research data from research institutions. 

    \item P4 used Ethics Pathways to paint a picture of power structures within the organization, discussing the moral dilemmas faced by a junior researcher when working on an unethical corporate project. Despite recognizing the company's unethical practices, the participant felt constrained by their position within the organizational hierarchy and struggled to refuse participation, which conflicted with their personal ethical standards. By going through the path, P4 reflected on moments of personal introspection and the influence of organizational structures on their decisions, steered by the questions from the facilitator. 

    \item P6 shared their journey of transparency and trust.  As an assistant professor collaborating with a senior professor on a project funded by a government agency, P6 found that undisclosed political motivations emerged at the end of the project. This led to frustration among research participants. Feeling powerless as the lead researcher on individual studies within the project, P6 highlighted structural issues stemming from disparities in stakeholder information and unclear communication. Despite these challenges, they recognized instances where clearer communication could have been initiated to address issues more effectively using the alternate pathways.

\end{itemize}

In our research, we encouraged participants to reflect on their past experiences as a way to construct in-depth narratives of engaging with the ethics flows. We aimed to capture the multifaceted layers involved in ethical decision-making. Echoing Durgahee's perspective on the value of reflection through storytelling \cite{durgahee1997reflective}, our approach enabled participants to critically recount their experiences and decisions, especially in situations with ethical nuances or sensitivity. throughout the journey described by the path highlights how engaging with ethics is a social process that often involves interactions among stakeholders with varying goals and different amounts of social power \cite{gray2019ethicalmediation,widder2024power}. Through this activity, participants could either resolve ethical dilemmas or gain deeper insight into the factors driving them through reflection.

Ethics Pathways helps us conceptualize ethics in a way that draws our attention to ethics flows in practice (rather than focusing on outcomes). Importantly, \textbf{our focus on flows opens a problem space, highlighting how research ethics are not solely decisions made by individual researchers but are ongoing practices that are embedded and situated in a broader set of interconnected affective experiences, institutional infrastructure, power structures, and social relationships.} The relationships are dynamic and ongoing and are constantly being enacted (or challenged). This also builds on some existing research, considering the affective experience and tacit skills required when dealing with research failures \cite{howell2021crackssuccess} or building relationships with community participants \cite{ledantec2015strangers}. We believe that this research suggests several new opportunities and directions for HCI and design researchers interested in design for research ethics:

\begin{itemize}
    \item \textit{Designing ethical changes and interventions in organizational processes}: Ethics engagement, in part, depends on the organizational and institutional context. What might it look like to consider ethics engagement at an organizational or institutional level rather than an individual decision-making level? Future work can further explore, probe, or seek to change factors like organizational policies, programs, and structures. Through this process, individuals are not only encouraged to reflect on their actions but also provided with opportunities to understand events from various perspectives, fostering a diversified approach to problem-solving.
    \item \textit{Designing to help researchers navigate power-laden contexts in practice}: Researchers' actions related to ethical engagement do not occur within a vacuum but in a complex knot \cite{jackson2014policyknot} of institutional systems and policies, as well as power-laden social interactions with other researchers and stakeholders. Future work can consider investigating tools, practices, or strategies that can help researchers navigate these social and political contexts, acknowledging these social power differentials.
    \item \textit{Design for ongoing reflection}: Formal ethics documentation such as an IRB often only occurs once at the beginning of the research process. Inspired by reflective design \cite{Sengers2005}, reflection in action \cite{schon2017reflective}, and our conceptualizing of ethics engagement as an ongoing, enacted set of practices, we might create new tools that help support researchers' reflections on their ethics engagements in an ongoing, long-term manner.
\end{itemize}



In our design activity, we recognized that ethics engagement transcends grand ideas, emphasizing instead the significance of how individual perspectives are actioned. Ethical engagement is the culmination of myriad small choices made within specific contexts, akin to navigating a constantly shifting path or flow. This flow or path is constantly changing and adapting, with twists and turns just like the situations we face. The choices we make along the way—taking a left turn, going back – are connected to how we feel about them and are influenced by the broader landscape we traverse—the rules, safety infrastructure, and fellow travelers (other stakeholders) on the journey. Embedded within this dynamic is the necessity for critical reflection, which serves as a compass to assess the fairness and justice of our decisions and reminds us of the weight of responsibility they carry. This reflective process not only informs our present actions but also reshapes the trajectory of ethical flows and our paths, enabling us to make more informed choices and fostering a diversified problem-solving approach that embraces varied perspectives.

 
\subsection{Considering Ethics Pathways in Relation to Other Design Activities and Future Applications}

We consider how Ethics Pathways builds on other design activities while providing new perspectives, and we suggest future uses and application areas for Ethics Pathways.

Many current design activities that consider ethics primarily focus on speculating about potential "downstream" ethical harms resulting from research and development, often utilizing speculative methods like design fiction (e.g., \cite{Nathan2008DIS, soden2019chi4evil, Sturdee21Consequences, lindley2017implications, ballard2019judgment}). These activities aim to explore the broader impacts of research and development outcomes. However, they often overlook addressing the questions around ethics embedded within the researchers' situated and lived experience of the research process itself, which Ethics Pathways highlights. Ethics Pathways also attempts to allow for speculation on alternate pasts by allowing and encouraging speculation on what could have been done in their specific situations by reflecting on what had transpired before. 

Other ethics-oriented speculation activities \cite{pater2022nohumans, fiesler2019ethicalconsiderations} have shown how design fiction and speculative design can be utilized to reflect on research ethics, particularly as a way to consider the ethical harms that can occur in research projects that are approved (or do not require review) by an IRB, particularly when using publicly available data. Like our work, these consider how ethics is an ongoing process beyond formal documentation. In addition to these types of ethical issues, Ethics Pathways also highlights how other social interactions and institutional dynamics in the research process can require ethical engagement. 

In addition, Gray et al. \cite{gray2022practitioner} proposed a scaffolding method to assist HCI practitioners in contemplating ethical dilemmas. These tools and activities serve as evidence of using design as a self-reflective tool; however, their focus primarily lies in creating forward-looking plans for practitioners rather than Ethics Pathways' focus on reflecting on past experiences and learning about other researchers' practices. During our sessions, participants are encouraged to reflect not just on their own actions but also on the larger institutional dynamics, which encompass social and physical infrastructures. By prompting contemplation of past events and the intricate workings within institutions, we aim to cultivate a deeper understanding that goes beyond individual or team behavior. This involves exploring the complex interplay of institutional dynamics and their influence on ethical decision-making.


\subsubsection{Future Applications of Ethics Pathways}
Overall, we find that Ethics Pathways is a resource-efficient research probe for understanding individuals' ethics engagement processes. It could be applicable in a broad spectrum of environments, particularly in HCI and design research. While our development of Ethics Pathways focused on participants currently doing HCI research in academia (although some had prior industry experience), we note that Ethics Pathways may also have applications beyond this context.

\textit{Practitioners studying research and design ethics} can leverage its potential as a probe, providing a deeper understanding of how individuals navigate ethical decisions in embodied and situated contexts. This aspect enables facilitators and researchers to delve into participants' narratives, extracting valuable insights into the intricate processes of ethical flows. Beyond research contexts, Ethics Pathways could be used to understand the experiences of HCI researchers and practitioners grappling with ethical considerations in technology development in industry. However, careful consideration of confidentiality and anonymity may be required if participants from the industry have concerns about nondisclosure agreements or potentially sharing proprietary information. We believe that our design and facilitation considerations surrounding obfuscation, abstraction, and anonymization would allow industry participants to share their experiences of social interactions, organizational barriers, and their felt emotions without violating any secrecy agreements. 

\textit{For helping researchers critically reflect on their own experiences}, the Ethics Pathways activity serves as a learning and reflection tool, fostering a deeper understanding of one's own ethics engagement processes. Several participants in our playtesting sessions had new insights into their own practices and potential alternatives they could have taken while participating in the activity. The activity could be adopted and integrated into educational contexts and used to aid in teaching students and new researchers how to think reflexively and critically about ethical issues that emerge during their research process. Also, the generated stories from participants can offer versatile educational applications as well, serving as scenarios for ethics education training, and as a complementary tool alongside others like the Ethical Disclaimer \cite{ethicsdisclaimer} and Dilemma Postcards \cite{Chivukula2021diss} activities.

\textit{For future facilitators of Ethics Pathways}, we note that we found that the facilitation of the activity is a nuanced process, requiring substantial effort from the facilitator. Despite its seemingly straightforward concept, the facilitator plays a pivotal role in guiding participants through reflective questions, ensuring a flexible and holistic exploration of ethical challenges. Prior research using design activities in workshop settings has highlighted that the design and parameters of the activities alone are not enough to create a generative or reflective space for participants; the facilitators' choices in language, prompts to participants, responses to participants' questions, and their own understanding of the social context also help create the conditions for participants' ability to successfully engage in the activity \cite{wong2021timelines,rosner2016outof,andersen2019magicmachine}. Our matching insights underscore the active role of the facilitator.

For instance, we encouraged participants to engage in self-reflection voluntarily while designing and reviewing their paths. As part of this process, we invited and reminded participants to express their thoughts, emotions, and reflections by placing path cards and describing their actions at each step. When participants created alternative paths, we asked for some reflective prompts to help participants draw connections between the fictional alternative paths and their real-life experiences. Through this approach, we prompt participants to explore and reflect on the factors that inspired them to consider alternative options. Future facilitators of Ethics Pathways should be conscious of the active role they can play in helping to prompt participants to reflect throughout the activity.

\section{Conclusion}
This paper introduces \textit{Ethics Pathways,} a design activity for examining the complexities of HCI and design researchers' ethics engagement during their research process. The activity is resource-efficient and guides participants through sharing personal reflections. By depicting past experiences as a set of paths (as well as speculative alternative paths of action), the activity can serve both as a way to gain insight into participants' past engagement with ethics and can help participants self-reflect on their past actions. Importantly, through reflections during our design and facilitation process, we conceptualized ethics engagement as \textit{ongoing practices}, rather than a set of outcomes. This conceptualization allowed us to begin understanding connections between individual affective experiences, social interactions across power differences, and institutional goals. We suggest that a focus on ethics engagement, its ongoing practices and flows opens up a new design space for creating changes within institutions and social practices related to research. The Ethics Pathways activity has potential future applications in various contexts, serving researchers, educators, and technology professionals, where it can be used to help enhance understanding of in-situ ethics and promote responsible and ethical research practices.

\bibliographystyle{ACM-Reference-Format}
\bibliography{Main}

\appendix

\begin{figure}
    \centering
    \includegraphics[width=0.5\textwidth]{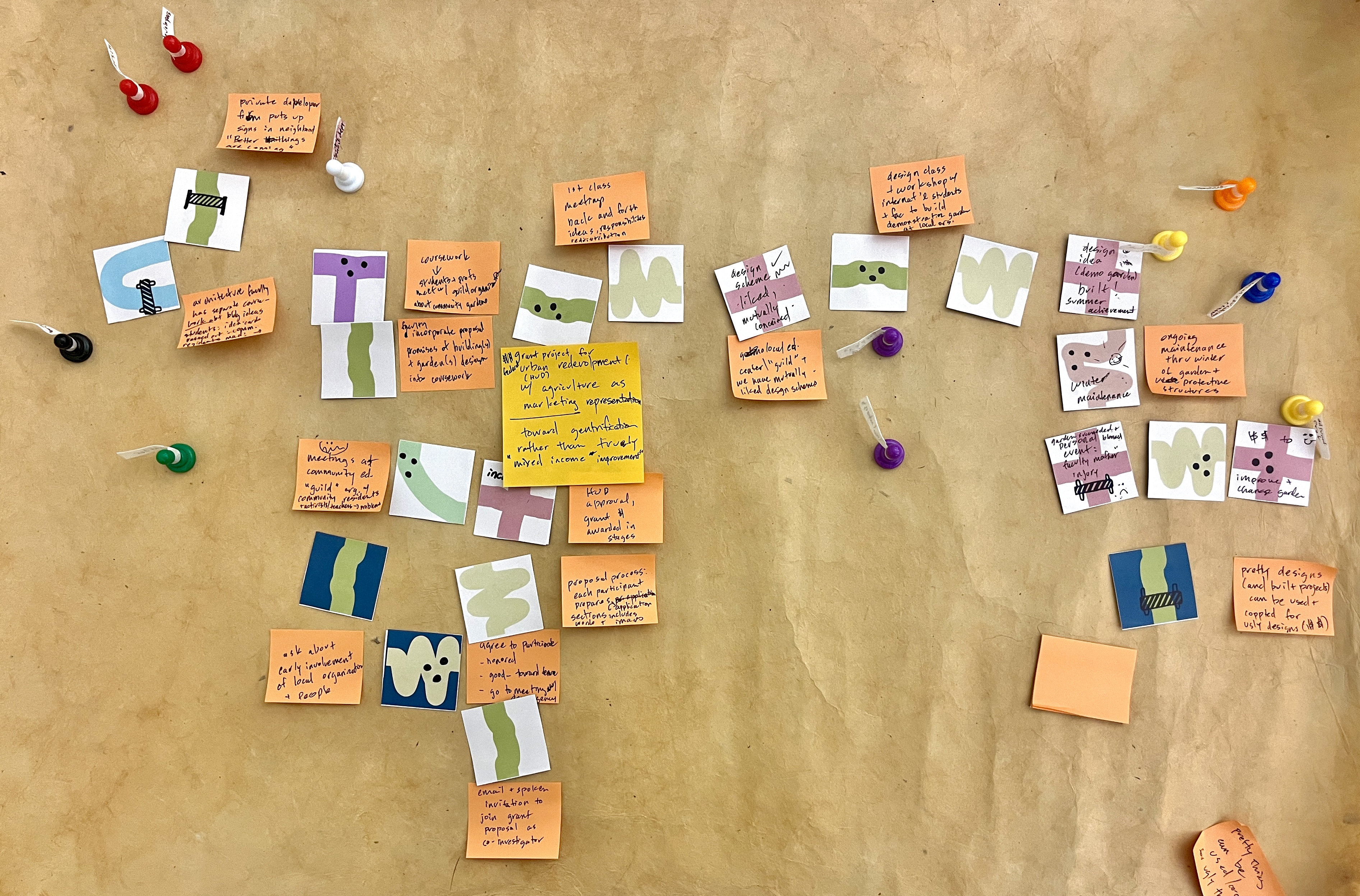} 
    \caption{P6's final results}
    \label{fig:p6map}
\end{figure}

\section{Narrative of P6 using Ethics Pathways}
\subsection{Incident}
P6 was discussing her involvement as a junior assistant professor in a grant project focused on the redevelopment of a historically lower-income neighborhood, particularly incorporating urban agriculture into the redevelopment plans. 

\subsection{Stakeholders}
Various partners were involved, including local community members, city agencies, educational institutions, an adjunct professor, an associate professor, local gardeners, developers, and P6 itself. 

\subsection{Journey}
P6 recounted their experience working on an urban planning grant project with a senior faculty member at a university. The project focused on the redevelopment of a historically lower-income neighborhood that had transitioned from an industrial area to a residential one. 

\begin{itemize}
    \item \textbf{P6 depicted her involvement as a straightforward path initially:} P6 received an email invitation to participate in a grant proposal. They agreed to join the project as co-investigators. P6 attended meetings and proceeded through the initial stages of the project smoothly. 

    \item \textbf{P6 depicted this process with a winding road card:} The proposal writing process commenced, involving collaboration and back-and-forth with stakeholders. P6 mentioned specific entities involved, such as the local education center and city development agencies. There were hierarchies and identities within the project stakeholders.

    \item \textbf{P6 used the road card with potholes:} P6's involvement mainly revolved around internal interactions during proposal preparation. P6 clarified that certain individuals were not directly involved in the project, emphasizing the role of the main investigator, who was the head of a local agency, and her own role as an assistant professor. P6 thought that the main investigator sought promotion and job advancement, which may have contributed to the project's visibility. There was significant press coverage during this phase, possibly due to the popularity of urban agriculture at the time. The grant project was eventually approved, marking a significant milestone in the process. 

    \item \textbf{P6 described barricades and holes in the process, with bridges representing progress and holes indicating potential problems:} The holes were particularly evident during discussions with the local community education center, highlighting issues such as a lack of community involvement in the project discussions. P6 and two other faculty members proposed ideas for gardens and buildings with agricultural or natural features to a local agency. After securing funding, they presented their ideas to the community, inviting input and participation. 

    \item \textbf{P6 designed the speculative paths regarding co-design with residents:} P6 questioned whether there should have been earlier involvement with the community before obtaining federal funding. They contemplated an alternative scenario where earlier community involvement could have led to better outcomes. P6 acknowledged that one of the faculty members had been actively engaged with the community for a long time, contrasting her own involvement primarily in university events. They suggested that earlier community engagement could have prevented potential problems or concerns raised by the community later in the process. 

    \item They divided responsibilities among themselves and others, including faculty members and local activists. P6 integrated the project into her classes, collaborating with students and community organizations. Classes were conducted to incorporate the project's themes into architecture, landscape architecture, and urban planning coursework. P6 referred to an upcoming community meeting where the project might be discussed further. P6 and others were involved in developing garden ideas and working with community members. 

    \item There was ongoing interaction and problem-solving through meetings and discussions. P6 encountered challenges with separate coursework and involvement in the project. There's a problem with community members getting the impression that all the proposed ideas for redesigning the community will be implemented, leading to negative associations with the project. 

    \item \textbf{P6 designed left-side paths to illustrate interactions with stakeholders who were left behind:}   Meanwhile, the redevelopment agency, in collaboration with private developers, is heavily involved in the project, aiming to profit from it by redeveloping the neighborhood, potentially leading to gentrification. Private developers are putting up signs in the neighborhood indicating impending development, despite the fact that they've forced people out of buildings they own with the expectation of receiving federal grant dollars to redevelop the area. 

    \item \textbf{Ending:} Despite the challenges and bumpy decisions, they continue working together, and a mutual design plan is developed, including a demonstration garden aimed at regenerating the polluted, post-industrial neighborhood. 
    
\end{itemize}

\subsection{Reflections}
P6 was concerned about how the project's success might have been used to further agendas tied to gentrification and racial dynamics in the community, even if that wasn't what they had intended. She had reflected on how well-meaning projects could end up being used for unintended purposes. P6 had also learned the importance of managing expectations and recognizing the potential for projects to be utilized in ways that did not align with their original intentions. This reflected a deeper understanding of the complexities involved in community projects.

\subsection{Thoughts and Opinions about Ethics Pathways}
She provided feedback on the facilitation method used, appreciating certain aspects, such as the gradual approach to information dissemination and the use of hand-drawn visuals. However, she also suggested improvements, such as the need for more pawns and considerations regarding color choices, indicating a thoughtful reflection on the process itself. 

\end{document}